\documentclass[11pt,oneside,reqno,american]{amsart}
\usepackage[T1]{fontenc}
\usepackage[latin9]{inputenc}
\usepackage{verbatim}
\usepackage{amstext}
\usepackage{amsthm}
\usepackage{amssymb}
\usepackage{xargs}[2008/03/08]

\makeatletter
\numberwithin{equation}{section}
\numberwithin{figure}{section}
\theoremstyle{plain}
\newtheorem{thm}{\protect\theoremname}[section]
\theoremstyle{definition}
\newtheorem{defn}[thm]{\protect\definitionname}
\theoremstyle{definition}
\newtheorem{example}[thm]{\protect\examplename}
\theoremstyle{remark}
\newtheorem{rem}[thm]{\protect\remarkname}
\theoremstyle{remark}
\newtheorem*{acknowledgement*}{\protect\acknowledgementname}

\usepackage{mathtools}
\usepackage{amsthm}
\usepackage{pdfsync} 
\usepackage{hyperref}
\usepackage[all]{xy}

 \theoremstyle{plain}
  
\usepackage{ETbb}
\usepackage[stix2,scaled=0.94]{newtxmath}%<<<<<< GOOD
\usepackage[cal=boondoxo,bb=boondox,frak=boondox]{mathalfa}%%%%%%% GOOD  EXcellent
\usepackage[scaled=.9]{rsfso}
\usepackage{xcolor} % Required for specifying colors by name
\definecolor{brown(traditional)}{rgb}{0.59, 0.29, 0.0}
\definecolor{blue(ryb)}{rgb}{0.01, 0.28, 1.0}
\definecolor{red}{rgb}{1.0, 0.0, 0.0}
\definecolor{magenta}{rgb}{1.0, 0.0, 1.0}
\definecolor{mahogany}{rgb}{0.75, 0.25, 0.0}
\definecolor{lavenderpurple}{rgb}{0.59, 0.48, 0.71}
\definecolor{olive}{rgb}{0.5, 0.5, 0.0}
\definecolor{brickred}{rgb}{0.8, 0.25, 0.33}
\definecolor{antiquefuchsia}{rgb}{0.57, 0.36, 0.51}
\definecolor{bole}{rgb}{0.47, 0.27, 0.23}
\definecolor{darkolivegreen}{rgb}{0.33, 0.42, 0.18}
\definecolor{deepjunglegreen}{rgb}{0.0, 0.29, 0.29}
\definecolor{brickred}{rgb}{0.8, 0.25, 0.33}
\definecolor{deepjunglegreen}{rgb}{0.0, 0.29, 0.29}
\definecolor{darkpastelgreen}{rgb}{0.01, 0.75, 0.24}
\definecolor{green(pigment)}{rgb}{0.0, 0.65, 0.31}
\definecolor{junglegreen}{rgb}{0.16, 0.67, 0.53}
\definecolor{officegreen}{rgb}{0.0, 0.5, 0.0}
\definecolor{seagreen}{rgb}{0.18, 0.55, 0.34}
\definecolor{teal}{rgb}{0.0, 0.5, 0.5}
\definecolor{brightgreen}{rgb}{0.4, 1.0, 0.0}
\definecolor{electricgreen}{rgb}{0.0, 1.0, 0.0}
\definecolor{malachite}{rgb}{0.04, 0.85, 0.32}

\usepackage{accents}
\makeatother

\usepackage{babel}
\providecommand{\acknowledgementname}{Acknowledgement}
\providecommand{\definitionname}{Definition}
\providecommand{\examplename}{Example}
\providecommand{\remarkname}{Remark}
\providecommand{\theoremname}{Theorem}

\begin{document}
\begin{comment}
Basic Math
\end{comment}

\global\long\def\ga{\alpha}%
\global\long\def\gb{\beta}%
\global\long\def\ggm{\gamma}%
\global\long\def\go{\omega}%
\global\long\def\gs{\sigma}%
\global\long\def\gd{\delta}%
\global\long\def\gD{\Delta}%
\global\long\def\vph{\phi}%
\global\long\def\gf{\phi}%
\global\long\def\gk{\kappa}%
\global\long\def\gl{\lambda}%
\global\long\def\gz{\zeta}%
\global\long\def\gh{\eta}%
\global\long\def\gy{\upsilon}%
\global\long\def\gth{\theta}%
\global\long\def\gO{\Omega}%
\global\long\def\gG{\Gamma}%

\global\long\def\eps{\varepsilon}%
\global\long\def\epss#1#2{\varepsilon_{#2}^{#1}}%
\global\long\def\ep#1{\eps_{#1}}%

\global\long\def\wh#1{\widehat{#1}}%
\global\long\def\hi{\hat{\imath}}%
\global\long\def\hj{\hat{\jmath}}%
\global\long\def\hk{\hat{k}}%
\global\long\def\ol#1{\overline{#1}}%
\global\long\def\ul#1{\underline{#1}}%

\global\long\def\spec#1{\textsf{#1}}%

\global\long\def\ui{\wh{\boldsymbol{\imath}}}%
\global\long\def\uj{\wh{\boldsymbol{\jmath}}}%
\global\long\def\uk{\widehat{\boldsymbol{k}}}%

\global\long\def\uI{\widehat{\mathbf{I}}}%
\global\long\def\uJ{\widehat{\mathbf{J}}}%
\global\long\def\uK{\widehat{\mathbf{K}}}%

\global\long\def\mc#1{\mathcal{#1}}%
\global\long\def\bs#1{\boldsymbol{#1}}%
\global\long\def\vect#1{\mathbf{#1}}%
\global\long\def\bi#1{\textbf{\emph{#1}}}%

\global\long\def\uv#1{\widehat{\boldsymbol{#1}}}%
\global\long\def\cross{\times}%

\global\long\def\di{d}%
\global\long\def\dee#1{\mathop{d#1}}%

\global\long\def\ddt{\frac{\dee{}}{\dee t}}%
\global\long\def\dbyd#1{\frac{\dee{}}{\dee{#1}}}%
\global\long\def\dby#1#2{\frac{\partial#1}{\partial#2}}%
\global\long\def\dxdt#1{\frac{\dee{#1}}{\dee t}}%

\global\long\def\vct#1{\bs{#1}}%

\begin{comment}
General Math
\end{comment}

\global\long\def\partialby#1#2{\frac{\partial#1}{\partial x^{#2}}}%
\newcommandx\parder[2][usedefault, addprefix=\global, 1=]{\frac{\partial#2}{\partial#1}}%
\global\long\def\supdot{^{\!\bs{\mathord{\cdot}}}}%

\global\long\def\fall{,\quad\text{for all}\quad}%

\global\long\def\reals{\mathbb{R}}%

\global\long\def\rthree{\reals^{3}}%
\global\long\def\rsix{\reals^{6}}%
\global\long\def\rn{\reals^{n}}%
\global\long\def\eucl{\mathbb{E}}%
\global\long\def\euthree{\eucl^{3}}%
\global\long\def\euln{\eucl^{n}}%

\global\long\def\prn{\reals^{n+}}%
\global\long\def\nrn{\reals^{n-}}%
\global\long\def\cprn{\overline{\reals}^{n+}}%
\global\long\def\cnrn{\overline{\reals}^{n-}}%
\global\long\def\rt#1{\reals^{#1}}%
\global\long\def\rtw{\reals^{12}}%

\global\long\def\les{\leqslant}%
\global\long\def\ges{\geqslant}%

\global\long\def\dX{\dee{\bp}}%
\global\long\def\dx{\dee x}%
\global\long\def\D{D}%

\global\long\def\from{\colon}%
\global\long\def\tto{\longrightarrow}%
\global\long\def\lmt{\longmapsto}%
\global\long\def\lhr{\lhook\joinrel\longrightarrow}%
\global\long\def\mto{\mapsto}%

\global\long\def\abs#1{\left|#1\right|}%

\global\long\def\isom{\cong}%

\global\long\def\comp{\circ}%

\global\long\def\cl#1{\overline{#1}}%

\global\long\def\fun{\varphi}%

\global\long\def\interior{\textrm{Int}\,}%
\global\long\def\inter#1{\kern0pt  #1^{\mathrm{o}}}%
\global\long\def\interior{\textrm{Int}\,}%
\global\long\def\inter#1{\kern0pt  #1^{\mathrm{o}}}%
\global\long\def\into{\mathrm{o}}%

\global\long\def\sign{\textrm{sign}\,}%
\global\long\def\sgn#1{(-1)^{#1}}%
\global\long\def\sgnp#1{(-1)^{\abs{#1}}}%

\global\long\def\du#1{#1^{*}}%

\global\long\def\tsum{{\textstyle \sum}}%
\global\long\def\lsum{{\textstyle \sum}}%

\global\long\def\dimension{\textrm{dim}\,}%

\global\long\def\esssup{\textrm{ess}\,\sup}%

\global\long\def\ess{\textrm{{ess}}}%

\global\long\def\kernel{\mathop{\textrm{\textup{Kernel}}}}%

\global\long\def\support{\mathop{\textrm{\textup{supp}}}}%

\global\long\def\image{\mathop{\textrm{\textup{Image}}}}%

\global\long\def\diver{\mathop{\textrm{\textup{div}}}}%

\global\long\def\spanv{\textrm{span}}%

\global\long\def\tr{\mathop{\textrm{\textup{tr}}}}%
\global\long\def\tran{\mathrm{tr}}%

\global\long\def\opt{\mathrm{opt}}%

\global\long\def\resto#1{|_{#1}}%
\global\long\def\incl{\mathcal{I}}%
\global\long\def\iden{\imath}%
\global\long\def\idnt{\textrm{Id}}%
\global\long\def\rest{\rho}%
\global\long\def\extnd{e_{0}}%

\global\long\def\proj{\textrm{pr}}%

\global\long\def\L#1{L\bigl(#1\bigr)}%
\global\long\def\LS#1{L_{S}\bigl(#1\bigr)}%

\global\long\def\ino#1{\int_{#1}}%

\global\long\def\half{\frac{1}{2}}%
\global\long\def\shalf{{\scriptstyle \half}}%
\global\long\def\third{\frac{1}{3}}%

\global\long\def\empt{\varnothing}%

\global\long\def\innp#1#2{\left\langle #1,#2\right\rangle }%

\global\long\def\paren#1{\left(#1\right)}%
\global\long\def\bigp#1{\bigl(#1\bigr)}%
\global\long\def\biggp#1{\biggl(#1\biggr)}%
\global\long\def\Bigp#1{\Bigl(#1\Bigr)}%

\global\long\def\braces#1{\left\{  #1\right\}  }%
\global\long\def\sqbr#1{\left[#1\right]}%
\global\long\def\anglep#1{\left\langle #1\right\rangle }%

\global\long\def\bigabs#1{\bigl|#1\bigr|}%
\global\long\def\dotp#1{#1^{\centerdot}}%
\global\long\def\pdot#1{#1^{\bs{\!\cdot}}}%

\global\long\def\eq{\sim}%
\global\long\def\quot{/\!\!\eq}%
\global\long\def\by{\!/\!}%

\begin{comment}
Symmetric Tensors and Multi-Indices
\end{comment}

\global\long\def\stp{\text{\small\ensuremath{\bigodot}}}%
\global\long\def\tp{\text{\small\ensuremath{\bigotimes}}}%

\global\long\def\mi#1{#1}%
\global\long\def\mii{I}%
\global\long\def\mie#1#2{#1_{1}\cdots#1_{#2}}%

\global\long\def\smi#1{\boldsymbol{#1}}%
\global\long\def\asmi#1{#1}%
\global\long\def\ordr#1{\left\langle #1\right\rangle }%
\begin{comment}
Indices and Multi-indices
\end{comment}

\global\long\def\symm#1{\paren{#1}}%
\global\long\def\smtr{\mathcal{S}}%

\global\long\def\perm{p}%
\global\long\def\sperm{\mathcal{P}}%

\begin{comment}
Lists
\end{comment}
\global\long\def\oneto{1,\dots,}%

\global\long\def\lisub#1#2#3{#1_{1}#2\dots#2#1_{#3}}%

\global\long\def\lisup#1#2#3{#1^{1}#2\dots#2#1^{#3}}%

\global\long\def\lisubb#1#2#3#4{#1_{#2}#3\dots#3#1_{#4}}%

\global\long\def\lisubbc#1#2#3#4{#1_{#2}#3\cdots#3#1_{#4}}%

\global\long\def\lisubbwout#1#2#3#4#5{#1_{#2}#3\dots#3\widehat{#1}_{#5}#3\dots#3#1_{#4}}%

\global\long\def\lisubc#1#2#3{#1_{1}#2\cdots#2#1_{#3}}%

\global\long\def\lisupc#1#2#3{#1^{1}#2\cdots#2#1^{#3}}%

\global\long\def\lisupp#1#2#3#4{#1^{#2}#3\dots#3#1^{#4}}%

\global\long\def\lisuppc#1#2#3#4{#1^{#2}#3\cdots#3#1^{#4}}%

\global\long\def\lisuppwout#1#2#3#4#5#6{#1^{#2}#3#4#3\wh{#1^{#6}}#3#4#3#1^{#5}}%

\global\long\def\lisubbwout#1#2#3#4#5#6{#1_{#2}#3#4#3\wh{#1}_{#6}#3#4#3#1_{#5}}%

\global\long\def\lisubwout#1#2#3#4{#1_{1}#2\dots#2\widehat{#1}_{#4}#2\dots#2#1_{#3}}%

\global\long\def\lisupwout#1#2#3#4{#1^{1}#2\dots#2\widehat{#1^{#4}}#2\dots#2#1^{#3}}%

\global\long\def\lisubwoutc#1#2#3#4{#1_{1}#2\cdots#2\widehat{#1}_{#4}#2\cdots#2#1_{#3}}%

\global\long\def\twp#1#2#3{\dee{#1}^{#2}\wedge\dee{#1}^{#3}}%

\global\long\def\thp#1#2#3#4{\dee{#1}^{#2}\wedge\dee{#1}^{#3}\wedge\dee{#1}^{#4}}%

\global\long\def\fop#1#2#3#4#5{\dee{#1}^{#2}\wedge\dee{#1}^{#3}\wedge\dee{#1}^{#4}\wedge\dee{#1}^{#5}}%

\global\long\def\idots#1{#1\dots#1}%
\global\long\def\icdots#1{#1\cdots#1}%

\global\long\def\norm#1{\|#1\|}%

\global\long\def\nonh{\heartsuit}%

\global\long\def\nhn#1{\norm{#1}^{\nonh}}%

\global\long\def\bigmid{\,\bigl|\,}%

\global\long\def\trps{^{{\scriptscriptstyle \textsf{T}}}}%

\global\long\def\testfuns{\mathcal{D}}%

\global\long\def\ntil#1{\tilde{#1}{}}%

\begin{comment}
Points Vectors and Regions
\end{comment}

\global\long\def\pis{y}%
\global\long\def\xo{\pis_{0}}%
\global\long\def\x{x}%

\global\long\def\pib{x}%
\global\long\def\bp{X}%
\global\long\def\ii{i}%
\global\long\def\ia{\alpha}%
\global\long\def\fp{y}%
\global\long\def\piv{v}%

\global\long\def\ib{i}%
\global\long\def\is{\alpha}%

\global\long\def\pbndo{\Gamma}%
\global\long\def\bndoo{\pbndo_{0}}%
 
\global\long\def\bndot{\pbndo_{t}}%
\global\long\def\intb{\inter{\body}}%
\global\long\def\bndb{\bdry\body}%

\global\long\def\cloo{\cl{\gO}}%

\global\long\def\nor{\nu}%
\global\long\def\Nor{\mathbf{N}}%

\global\long\def\dA{\dee A}%

\global\long\def\dV{\dee V}%

\global\long\def\eps{\varepsilon}%

\global\long\def\tv{v}%
\global\long\def\av{u}%

\global\long\def\svs{\mathcal{W}}%
\global\long\def\vs{\mathbf{V}}%
\global\long\def\avs{\mathbf{U}}%
\global\long\def\affsp{\mathcal{A}}%
\global\long\def\man{\mathcal{M}}%
\global\long\def\odman{\mathcal{N}}%
\global\long\def\subman{\mathcal{V}}%
\global\long\def\pt{p}%

\global\long\def\vbase{e}%
\global\long\def\sbase{\mathbf{e}}%
\global\long\def\msbase{\mathfrak{e}}%
\global\long\def\vect{v}%
\global\long\def\dbase{\sbase}%

\begin{comment}
{*}{*}{*}{*}{*} Jets and Vector Bundles {*}{*}{*}{*}{*}
\end{comment}

\global\long\def\chart{\varphi}%
\global\long\def\Chart{\Phi}%

\global\long\def\mind{\alpha}%
\global\long\def\vb{W}%
\global\long\def\vbp{\pi}%

\global\long\def\vbt{\mathcal{E}}%
\global\long\def\fib{\vs}%
\global\long\def\vbts{W}%
\global\long\def\avb{U}%
\global\long\def\vbp{\xi}%

\global\long\def\chart{\vph}%
\global\long\def\vbchart{\Phi}%

\global\long\def\jetb#1{J^{#1}}%
\global\long\def\jet#1{j^{1}(#1)}%
\global\long\def\tjet{\tilde{\jmath}}%

\global\long\def\Jet#1{J^{1}(#1)}%

\global\long\def\jetm{j}%

\global\long\def\coj{\mathfrak{d}}%

\begin{comment}
Forms-Differential Geometry
\end{comment}

\global\long\def\alt{\mathfrak{A}}%

\global\long\def\pou{\eta}%

\global\long\def\ext{{\textstyle \bigwedge}}%
\global\long\def\forms{\Omega}%

\global\long\def\dotwedge{\dot{\mbox{\ensuremath{\wedge}}}}%

\global\long\def\vel{\theta}%
\begin{comment}
<volume element
\end{comment}

\global\long\def\Jac{\mathcal{J}}%

\global\long\def\contr{\mathbin{\raisebox{0.4pt}{\mbox{\ensuremath{\lrcorner}}}}}%
\global\long\def\fcor{\llcorner}%
\global\long\def\bcor{\lrcorner}%
\global\long\def\fcontr{\mathbin{\raisebox{0.4pt}{\mbox{\ensuremath{\llcorner}}}}}%

\global\long\def\lie{\mathcal{L}}%

\global\long\def\ssym#1#2{\ext^{#1}T^{*}#2}%

\global\long\def\sh{^{\sharp}}%

\global\long\def\nfo{\ext^{n}T^{*}\base}%
\global\long\def\dfs{\ext^{d}T^{*}\base}%
\global\long\def\dmfs{\ext^{d-1}T^{*}\base}%

\begin{comment}
>\textcompwordmark >\textcompwordmark >\textcompwordmark >\textcompwordmark >\textcompwordmark >Bodies
Space Time Events<\textcompwordmark <\textcompwordmark <\textcompwordmark <\textcompwordmark <\textcompwordmark <\textcompwordmark <\textcompwordmark <
\end{comment}

\global\long\def\spc{\mathcal{S}}%
\global\long\def\sptm{\mathcal{E}}%
\global\long\def\evnt{e}%
\global\long\def\frame{\Psi}%

\global\long\def\timeman{\mathcal{T}}%
\global\long\def\zman{t}%
\global\long\def\dims{n}%
\global\long\def\m{\dims-1}%
\global\long\def\dimw{m}%

\global\long\def\wc{z}%

\global\long\def\fourv#1{\mbox{\ensuremath{\mathfrak{#1}}}}%

\global\long\def\body{\mathcal{B}}%
\global\long\def\man{\mathcal{M}}%
\global\long\def\var{\mathcal{V}}%
\global\long\def\base{\mathcal{X}}%
\global\long\def\fb{\mathcal{Y}}%
\global\long\def\srfc{\mathcal{Z}}%
\global\long\def\dimb{n}%
\global\long\def\dimf{m}%
\global\long\def\afb{\mathcal{Z}}%

\global\long\def\bdry{\partial}%

\global\long\def\gO{\varOmega}%

\global\long\def\reg{\gO}%
\global\long\def\bdrr{\bdry\reg}%

\global\long\def\bdom{\bdry\gO}%

\global\long\def\bndo{\partial\gO}%

\global\long\def\tpr{\vartheta}%

\begin{comment}
Kinematics, Configurations, Strains
\end{comment}

\global\long\def\mot{M}%
\global\long\def\vf{w}%
\global\long\def\const{h}%

\global\long\def\avf{u}%

\global\long\def\stn{\varepsilon}%
\global\long\def\djet{\chi}%

\global\long\def\jvf{\eps}%

\global\long\def\rig{r}%

\global\long\def\rigs{\mathcal{R}}%

\global\long\def\qrigs{\!/\!\rigs}%

\global\long\def\qd{\!/\,\!\kernel\diffop}%

\global\long\def\dis{\chi}%
\global\long\def\conf{\kappa}%
\global\long\def\invc{\hat{\conf}^{-1}}%
\global\long\def\dinvc{\hat{\conf}^{-1*}}%
\global\long\def\csp{\mathcal{Q}}%

\global\long\def\embds{\textrm{Emb}}%

\global\long\def\lc{A}%

\global\long\def\lv{\dot{A}}%
\global\long\def\alv{\dot{B}}%

\begin{comment}
Jets:
\end{comment}

\global\long\def\j{\mathop{\mathrm{j}}}%
\global\long\def\mapp{M}%
\global\long\def\J{J}%
\global\long\def\jex{\mathop{}\!\mathrm{j}}%

\begin{comment}
\%\%\%\%\%\%\%\%\%\%\%\%\%\%\%\%\%\%\%\%\%\%\%\%\%\%\%\%\%\%\%\%

Forces and Stresses

\%\%\%\%\%\%\%\%\%\%\%\%\%\%\%\%\%\%\%\%\%\%\%\%\%\%\%\%\%\%\%
\end{comment}

\global\long\def\fc{F}%
\global\long\def\load{f}%
\global\long\def\afc{g}%

\global\long\def\bfc{\mathbf{b}}%
\global\long\def\bfcc{b}%

\global\long\def\sfc{\mathbf{t}}%
\global\long\def\sfcc{t}%

\begin{comment}
Stresses
\end{comment}

\global\long\def\stm{\varsigma}%
\global\long\def\std{S}%
\global\long\def\tst{\sigma}%
\global\long\def\tstd{s}%
\global\long\def\st{\sigma}%
\global\long\def\vst{\varsigma}%
\global\long\def\vstd{S}%
\global\long\def\tstm{\sigma}%
\global\long\def\vstm{\varsigma}%
\begin{comment}
<\textcompwordmark < Variational stress
\end{comment}

\global\long\def\stp{S_{P}}%
\global\long\def\slf{R}%
\begin{comment}
Self force and stress principal components
\end{comment}

\global\long\def\crel{\Phi}%
\begin{comment}
Constitutive relation as a section defined on the jet bundle
\end{comment}

\global\long\def\stmat{\tau}%

\global\long\def\gdiv{\bdry\textrm{iv\,}}%
\global\long\def\extjet{\mathfrak{d}}%

\global\long\def\smc#1{\mathfrak{#1}}%

\global\long\def\nhs{P}%
\begin{comment}
Nonholonomic Variational Second Order Stress
\end{comment}
\global\long\def\nhsa{P}%
\global\long\def\nhsb{\underline{P}}%
\begin{comment}
<\textcompwordmark <\textcompwordmark <\textcompwordmark < Components
of nonholonomic stresses
\end{comment}

\global\long\def\soc{Z}%
\begin{comment}
Second Order Cauchy Stress
\end{comment}

\global\long\def\sts{\varSigma}%
\begin{comment}
spaces of stresses 
\end{comment}
\global\long\def\spstd{\mathfrak{S}}%
\global\long\def\sptst{\mathfrak{T}}%
\global\long\def\spnhs{\mathcal{P}}%
\global\long\def\Ljj{\L{J^{1}(J^{k-1}\vb),\ext^{n}T^{*}\base}}%

\global\long\def\spsb{\text{\Large\ensuremath{\Delta}}}%

\global\long\def\ened{\mathfrak{w}}%
\global\long\def\energy{\mathfrak{W}}%

\global\long\def\ebdfc{T}%
\global\long\def\optimum{\st^{\textrm{opt}}}%
\global\long\def\scf{K}%

\begin{comment}
\%\%\%\%\%\%\%\%\%\%\%\%\%\%\%\%\%\%\%\%

\%\%\%\% Group Action \%\%\%\%\%\%\%\%\%

\%\%\%\%\%\%\%\%\%\%\%\%\%\%\%\%\%\%\%\%
\end{comment}

\global\long\def\grp{G}%
\global\long\def\gact{A}%
\global\long\def\gid{e}%
\global\long\def\gel{\ggm}%

\global\long\def\ael{\upsilon}%
\global\long\def\lal{\mathfrak{g}}%

\begin{comment}
{*}{*}{*}{*}{*}{*}{*}{*}{*}{*}{*}{*}{*}{*}{*}{*}{*}{*}{*}{*}{*}{*}{*}{*}{*}{*}{*}{*}{*}{*}{*}{*}{*}{*}{*}{*}{*}{*}{*}{*}{*}{*}

{*}{*}{*}{*}{*}{*}{*}{*}{*}{*}{*}{*}{*}{*}{*}{*}{*}{*}{*}Cauchy Fluxes{*}{*}{*}{*}{*}{*}{*}{*}{*}{*}

{*}{*}{*}{*}{*}{*}{*}{*}{*}{*}{*}{*}{*}{*}{*}{*}{*}{*}{*}{*}{*}{*}{*}{*}{*}{*}{*}{*}{*}{*}{*}{*}{*}{*}{*}{*}{*}{*}{*}{*}{*}{*}{*}
\end{comment}

\global\long\def\prop{P}%
\global\long\def\expr{\Pi}%

\global\long\def\aprop{Q}%

\global\long\def\flux{\omega}%
\global\long\def\aflux{\psi}%

\global\long\def\fform{\tau}%

\global\long\def\dimn{n}%

\global\long\def\sdim{{\dimn-1}}%

\global\long\def\fdens{\phi}%

\global\long\def\pform{s}%
\global\long\def\vform{\beta}%
\global\long\def\sform{\tau}%
\global\long\def\flow{\vf}%
\global\long\def\n{\m}%
\global\long\def\cmap{\mathfrak{t}}%
\global\long\def\vcmap{\varSigma}%

\global\long\def\mvec{\mathfrak{v}}%
\global\long\def\mveco#1{\mathfrak{#1}}%
\global\long\def\mv#1{\mathfrak{#1}}%
\global\long\def\smbase{\mathfrak{e}}%
\global\long\def\spx{\simp}%
\global\long\def\il{l}%
\global\long\def\awe{\frown}%
\begin{comment}
<\textcompwordmark <\textcompwordmark <\textcompwordmark < multivectors
\end{comment}

\global\long\def\hp{H}%
\global\long\def\ohp{h}%

\global\long\def\hps{G_{\dims-1}(T\spc)}%
\global\long\def\ohps{G_{\dims-1}^{\perp}(T\spc)}%

\global\long\def\hyper{\mathcal{S}}%

\global\long\def\hpsx{G_{\dims-1}(\tspc)}%
\global\long\def\ohpsx{G_{\dims-1}^{\perp}(\tspc)}%

\global\long\def\fbun{F}%

\global\long\def\flowm{\Phi}%

\global\long\def\tgb{T\spc}%
\global\long\def\ctgb{T^{*}\spc}%
\global\long\def\tspc{T_{\pis}\spc}%
\global\long\def\dspc{T_{\pis}^{*}\spc}%

\begin{comment}
{*}{*}{*}{*}{*} ELECTROMAGNETISM IN SPACETIME <\textcompwordmark <\textcompwordmark <\textcompwordmark <\textcompwordmark <\textcompwordmark <
\end{comment}

%%%%%%% ELECTROMAGNETISM IN SPACETIME %%%%%%

\global\long\def\fflow{\fourv J}%
%      four-flow
\global\long\def\fvform{\mathfrak{b}}%
%      four body flux
\global\long\def\fsform{\mathfrak{t}}%
%      four surface flux
\global\long\def\fpform{\mathfrak{s}}%
%      Four production rate
%\newcommand{\fgr}{\mathfrak{I}}%         Growth rate in space time
\global\long\def\lfc{\mathfrak{F}}%

\global\long\def\maxw{\mathfrak{g}}%
%        Maxwell form or Stream form
\global\long\def\frdy{\mathfrak{f}}%
%        Faraday 2-form
%\newcommand{\maxw}{\mathcal{G}}%        Maxwell form or Stream form
%\newcommand{\frdy}{\mathcal{F}}%        Faraday 2-form
\global\long\def\ptnl{\gf}%
\global\long\def\pts{\Psi}%
\global\long\def\tptn{\Psi}%
\global\long\def\vptn{\mathfrak{a}}%
%                   Potential 1-form
\global\long\def\mtst{\tstd_{M}}%
\global\long\def\mvst{\vstd_{M}}%
%%%%%%%%%%%%%%%%%%%%%%%%%%%%%%%%%%%%%%%%%%%%%

\begin{comment}
Sobolev Spaces
\end{comment}

\global\long\def\sobp#1#2{W_{#2}^{#1}}%

\global\long\def\inner#1#2{\left\langle #1,#2\right\rangle }%

\global\long\def\fields{\sobp pk(\vb)}%

\global\long\def\bodyfields{\sobp p{k_{\partial}}(\vb)}%

\global\long\def\forces{\sobp pk(\vb)^{*}}%

\global\long\def\bfields{\sobp p{k_{\partial}}(\vb\resto{\bndo})}%

\global\long\def\loadp{(\sfc,\bfc)}%

\global\long\def\strains{\lp p(\jetb k(\vb))}%

\global\long\def\stresses{\lp{p'}(\jetb k(\vb)^{*})}%

\global\long\def\diffop{D}%

\global\long\def\strainm{E}%

\global\long\def\incomps{\vbts_{\yieldf}}%

\global\long\def\devs{L^{p'}(\eta_{1}^{*})}%

\global\long\def\incompsns{L^{p}(\eta_{1})}%

\begin{comment}
Distributions and Currents
\end{comment}

\global\long\def\testf{\mathcal{D}}%
\global\long\def\dists{\mathcal{D}'}%

\global\long\def\codiv{\boldsymbol{\partial}}%

\global\long\def\currof#1{\tilde{#1}}%

\global\long\def\chn{c}%
\global\long\def\chnsp{\mathbf{C}}%

\global\long\def\current{T}%
\global\long\def\curr{R}%

\global\long\def\curd{S}%
\global\long\def\curwd#1{\wh{#1}}%
\global\long\def\curnd#1{\wh{#1}}%

\global\long\def\contrf{{\scriptstyle \smallfrown}}%

\global\long\def\prodf{{\scriptstyle \smallsmile}}%

\global\long\def\form{\omega}%

\global\long\def\dens{\rho}%

\global\long\def\simp{s}%
\global\long\def\ssimp{\Delta}%
\global\long\def\cpx{K}%

\global\long\def\cell{C}%

\global\long\def\chain{B}%
\global\long\def\A{A}%
\global\long\def\B{B}%

\global\long\def\ach{A}%

\global\long\def\coch{X}%

\global\long\def\scale{s}%

\global\long\def\fnorm#1{\norm{#1}^{\flat}}%

\global\long\def\chains{\mathcal{A}}%

\global\long\def\ivs{\boldsymbol{U}}%

\global\long\def\mvs{\boldsymbol{V}}%

\global\long\def\cvs{\boldsymbol{W}}%

\global\long\def\ndual#1{#1'}%

\global\long\def\nd{'}%

\begin{comment}
Function Spaces
\end{comment}

\global\long\def\cee#1{C^{#1}}%

\global\long\def\lone{\{L^{1}\}}%

\global\long\def\linf{L^{\infty}}%

\global\long\def\lp#1{L^{#1}}%

\global\long\def\ofbdo{(\bndo)}%

\global\long\def\ofclo{(\cloo)}%

\global\long\def\vono{(\gO,\rthree)}%

\global\long\def\lomu{\{L^{1,\mu}\}}%
\global\long\def\limu{L^{\infty,\mu}}%
\global\long\def\limub{\limu(\body,\rthree)}%
\global\long\def\lomub{\lomu(\body,\rthree)}%

\global\long\def\vonbdo{(\bndo,\rthree)}%
\global\long\def\vonbdoo{(\bndoo,\rthree)}%
\global\long\def\vonbdot{(\bndot,\rthree)}%

\global\long\def\vonclo{(\cl{\gO},\rthree)}%

\global\long\def\strono{(\gO,\reals^{6})}%

\global\long\def\sob{\{W_{1}^{1}\}}%

\global\long\def\sobb{\sob(\gO,\rthree)}%

\global\long\def\lob{\lone(\gO,\rthree)}%

\global\long\def\lib{\linf(\gO,\reals^{12})}%

\global\long\def\ofO{(\gO)}%

\global\long\def\oneo{{1,\gO}}%
\global\long\def\onebdo{{1,\bndo}}%
\global\long\def\info{{\infty,\gO}}%

\global\long\def\infclo{{\infty,\cloo}}%

\global\long\def\infbdo{{\infty,\bndo}}%
\global\long\def\lobdry{\lone(\bdry\gO,\rthree)}%

\global\long\def\ld{LD}%

\global\long\def\ldo{\ld\ofO}%
\global\long\def\ldoo{\ldo_{0}}%

\global\long\def\trace{\gamma}%
\global\long\def\dtrace{\delta}%
\global\long\def\gtrace{\beta}%

\global\long\def\pr{\proj_{\rigs}}%

\global\long\def\pq{\proj}%

\global\long\def\qr{\,/\,\reals}%

\begin{comment}
Plasticity and Optimization
\end{comment}

\global\long\def\aro{S_{1}}%
\global\long\def\art{S_{2}}%

\global\long\def\mo{m_{1}}%
\global\long\def\mt{m_{2}}%

\global\long\def\ebdfc{T}%

\global\long\def\mini{\Omega}%
\global\long\def\optimum{s^{\mathrm{opt}}}%
\global\long\def\scf{K}%
\global\long\def\opsf{\st^{\mathrm{opt}}}%
\global\long\def\doptimum{s^{\opt,{\scriptscriptstyle D}}}%
\global\long\def\loptimum{s^{\opt,{\scriptscriptstyle \mathcal{M}}}}%

\global\long\def\fsubs{M}%

\begin{comment}
Optimization
\end{comment}

\global\long\def\yieldc{B}%

\global\long\def\yieldf{Y}%

\global\long\def\trpr{\pi_{P}}%

\global\long\def\devpr{\pi_{\devsp}}%

\global\long\def\prsp{P}%

\global\long\def\devsp{D}%

\global\long\def\ynorm#1{\|#1\|_{\yieldf}}%

\global\long\def\colls{\Psi}%
%Collapse sufrace 

\global\long\def\aro{S_{1}}%
\global\long\def\art{S_{2}}%

\global\long\def\mo{m_{1}}%
\global\long\def\mt{m_{2}}%

\global\long\def\trps{^{\mathsf{T}}}%

\global\long\def\hb{^{\mathrm{hb}}}%

\global\long\def\yieldst{s_{Y}}%

\global\long\def\yieldc{B}%

\global\long\def\lcap{C}%

\global\long\def\yieldf{Y}%

\global\long\def\sphpr{\pi_{P}}%

\global\long\def\devpr{\pi_{\devsp}}%

\global\long\def\prsp{P}%

\global\long\def\devsp{D}%

\global\long\def\ynorm#1{\|#1\|_{\yieldf}}%

\global\long\def\colls{\Psi}%
%Collapse sufrace 

\global\long\def\cone{Q}%
\global\long\def\fpr{\Pi}%
\global\long\def\fprd{\fpr_{\devsp}}%
\global\long\def\fprp{\fpr_{\prsp}}%
\global\long\def\find{I_{\devsp}}%
\global\long\def\finp{I_{\prsp}}%
\global\long\def\fnorm#1{\norm{#1}_{\devsp}}%

\begin{comment}
Invariant Stress Concentration
\end{comment}

\global\long\def\rig{r}%
\global\long\def\rigs{\mathcal{R}}%
\global\long\def\qrigs{\!/\!\rigs}%
\global\long\def\anv{\omega}%
\global\long\def\I{I}%
\global\long\def\mone{M_{1}}%

\global\long\def\bd{BD}%

\global\long\def\po{\proj_{0}}%
\global\long\def\normp#1{\norm{#1}'_{\ld}}%

\global\long\def\ssx{S}%

\global\long\def\smap{s}%

\global\long\def\smat{\chi}%

\global\long\def\sx{e}%

\global\long\def\snode{P}%
\global\long\def\newmacroname{}%

\global\long\def\elem{e}%

\global\long\def\nel{L}%

\global\long\def\el{l}%

\global\long\def\gr{g}%
\global\long\def\ngr{G}%

\global\long\def\eldof{\alpha}%

\global\long\def\glbs{\psi}%

\global\long\def\ipln{\phi}%

\global\long\def\ndof{D}%

\global\long\def\dof{d}%

\global\long\def\nldof{N}%

\global\long\def\ldof{n}%

\global\long\def\lvf{\chi}%

\global\long\def\amat{A}%
\global\long\def\bmat{B}%

\global\long\def\subsp{\mathcal{M}}%
\global\long\def\zerofn{Z}%

\global\long\def\snomat{E}%

\global\long\def\femat{E}%

\global\long\def\tmat{T}%

\global\long\def\fvec{f}%

\global\long\def\snsp{\mathcal{S}}%

\global\long\def\slnsp{\Phi}%
\global\long\def\dslnsp{\Phi^{{\scriptscriptstyle D}}}%

\global\long\def\ro{r_{1}}%

\global\long\def\rtwo{r_{2}}%

\global\long\def\rth{r_{3}}%

\global\long\def\fmax{M}%

\begin{comment}
{*}{*}{*}{*}{*}{*}{*}{*} Defects {*}{*}{*}{*}{*}{*}{*}{*}{*}{*}
\end{comment}

\global\long\def\dform{\psi}%
\begin{comment}
Defect Form
\end{comment}

\global\long\def\srfc{\mathcal{S}}%

\begin{comment}
Specific to Current Paper
\end{comment}

\global\long\def\semib{\mathrm{SB}}%

\global\long\def\tm#1{\overrightarrow{#1}}%
\global\long\def\tmm#1{\underrightarrow{\overrightarrow{#1}}}%

\global\long\def\itm#1{\overleftarrow{#1}}%
\global\long\def\itmm#1{\underleftarrow{\overleftarrow{#1}}}%

\global\long\def\ptrac{\mathcal{P}}%

\begin{comment}
Non-Holonomic Jets
\end{comment}

\global\long\def\nh#1{\hat{#1}}%
\global\long\def\nj{\hat{\jmath}}%
\global\long\def\nJ{\hat{J}}%
\global\long\def\rin#1{\mathfrak{#1}}%
\global\long\def\npi{\hat{\pi}}%
\global\long\def\rp{\rin p}%
\global\long\def\rq{\rin q}%
\global\long\def\rr{\rin r}%

\global\long\def\xty{(\base,\fb)}%
\global\long\def\xts{(\base,\spc)}%
\global\long\def\r{r}%
\global\long\def\ntm{(\reals^{n},\reals^{m})}%

\begin{comment}
Specific to this paper
\end{comment}

\global\long\def\tproj{\frame_{\timeman}}%
\global\long\def\sproj{\frame_{\spc}}%

\begin{comment}
Optimal velocity fields
\end{comment}

\global\long\def\cons{c}%
\global\long\def\optm{\go}%
\global\long\def\flxs{\mathcal{W}}%
\global\long\def\cost{Q}%

\begin{comment}
Growing bodies
\end{comment}

\global\long\def\mtn{e}%
\global\long\def\sppp{\lambda}%

\global\long\def\mtsp{\mathscr{E}}%

\global\long\def\disp{g}%
\global\long\def\diffs{G}%

\global\long\def\bv{BV}%

\begin{comment}
Charges and Multipoles
\end{comment}

\global\long\def\Charge{Q}%

\global\long\def\pole{q}%

\global\long\def\pdens{\rho}%

\global\long\def\ms#1{\mathfrak{#1}}%

\global\long\def\Hfl{\Phi}%

\title[Multipole Distributions]{\textsf{Multipole Distributions and Hyper-Flux Fields}}
\author{Vladimir Gol'dshtein$\vphantom{N^{2}}^{1}$  and Reuven Segev$\vphantom{N^{2}}^{2}$}
\address{}
\begin{abstract}
We outline here a simple mathematical introduction to the notions
of multipoles for a general extensive property $\expr$ from the
point of view of continuum mechanics. Classically, $\expr$ is the
electric charge, but the theory is not limited to electrostatics.
The proposed framework allows a simple computation of the bound ``charges''
and bound multipoles of lower orders. In addition, if the property
$\expr$ has a potential function in the sense described below, a
general expression for the mechanical force (power) functional acting
on bodies containing the property is presented. Finally, using a similar
viewpoint, we consider hyper-fluxes\textemdash flux fields of tensorial
order greater than one\textemdash and show that moving multipoles
(in particular, a moving dielectric) give rise to hyper-fluxes.
\end{abstract}

\date{\today\\[2mm]
$^1$ Department of Mathematics, Ben-Gurion University of the Negev, Israel. Email: vladimir@bgu.ac.il\\
$^2$ Department of Mechanical Engineering, Ben-Gurion University of the Negev, Israel. Email: rsegev@post.bgu.ac.il}

\maketitle

\section{Introduction}

We outline here a simple mathematical introduction to the notions
of multipoles for a general extensive property $\expr$ from the
point of view of continuum mechanics. Classically, $\expr$ is the
electric charge, but the theory is not limited to electrostatics.
The proposed framework allows a simple computation of the bound ``charges''
and bound multipoles of lower orders. In addition, if the property
$\expr$ has a potential function in the sense described below, a
general expression for the mechanical force (power) functional acting
on bodies containing the property is presented. Finally, using a similar
viewpoint, we consider hyper-fluxes\textemdash flux fields of tensorial
order greater than one\textemdash and show that moving multipoles
(in particular, a moving dielectric) give rise to hyper-fluxes.

Multipoles for the electrostatic field are considered, for example,
in \cite{Kafadar71,LandauLifshitz,Torres_etc_Multipoles,Jackson,Raab_Lange,Zangwill}.
They are presented as a means for approximating the electric potential
field,
\begin{equation}
\ptnl(x)=\int\frac{\rho(x')}{\abs{x-x'}}\dee{V(x')},\label{eq:potential}
\end{equation}
at point $x$ far from the charge with density $\rho$ that induces
it. Special attention is given to dipoles and continuous distributions
of dipoles in dielectrics. See, for example, the extended motivation
for the introduction of the polarization field in \cite[Chapter 6]{Zangwill}
and that of \cite[Chapte 2]{LandauLifshitz84}.

Forces on dialectics are derived in \cite{Stratton41,Panofsky1962,Penfield_Haus,Maugin80,LandauLifshitz84,Zangwill,Dorfamand_and_Ogden_2014}.
The total force and couple that act on a quadrupole are derived in
\cite[p.~104]{Zangwill}. Expressions for the total forces on discrete
multipoles in general are presented in \cite{Jackson,Raab_Lange}.

As mentioned above, the physical theory is based on the assumption
that there is some extensive property $\expr$, e.g., the electric
charge, for which there is a potential field $\ptnl$. We do not assume
any particular relation, such as (\ref{eq:potential}), between the
distribution of the property $\expr$ in space and the potential field
$\ptnl$. We view such a relation as a constitutive property. Thus,
we admit variations of the potential field while keeping the distribution
of the property fixed and vice versa. Modeling the physical space,
$\spc$, by $\rthree$, a potential field, or a variation thereof,
is assumed to be a smooth real-valued function of compact support\textemdash a
test function.

A distribution of an $r$-multipolar medium in space is a Schwartz
distribution, $\Charge$ specified by a given collection of Borel
measures $\{\pole^{i_{1}\dots i_{k}}\}$, $0\les k\les r$, in the
form
\begin{equation}
\Charge(\ptnl)=\sum_{0\les k\les r}\int_{\rthree}\ptnl_{,i_{1}\dots i_{k}}\dee{\pole^{i_{1}\dots i_{k}}},\label{eq:definition-1}
\end{equation}
where the summation convention is used, and a comma subscript indicates
partial differentiation.

Note that at this point, $\ptnl$ need not have any physical interpretation.
When $\ptnl$ is interpreted as a variation of the potential field,
$\Charge(\ptnl)$ is interpreted as the resulting variation of potential
energy.

The representation by measures allows the description of a concentrated
charge or a single dipole, as well as smooth distributions of multipoles.
In addition, the representation by measures enables the restriction
of the functional $\Charge$ to Borel measurable subsets $\body$
to obtain functionals
\begin{equation}
\Charge_{\body}(\gf):=\sum_{0\les k\les r}\int_{\body}\gf_{,i_{1}\dots i_{k}}\dee{\pole^{i_{1}\dots i_{k}}}.
\end{equation}
Thus, multipolar distribution may be restricted to measurable subsets.

For example, in Section \ref{sec:Bound-Charge-qudrupoles}, assuming
that the subset $\body$ is sufficiently regular so that the Green-Gauss
theorem is applicable, the bound charge and the bound polarization
induced by a smooth quadrupole distribution are computed. In particular,
a charge density per unit length is induced along the edges in $\bdry\body$.

In Section \ref{sec:Forces}, for simplicity, we identify a material
body with its image in the physical space under some reference configuration.
We view a mechanical force on the body as a bounded linear functional\textemdash force,
or power, functional\textemdash on the vector space of velocity fields
defined on the body. A motion of a body is modeled by a one-parameter
group of diffeomorphisms of $\rthree$, and it is assumed that the
multipolar distribution is carried with the body. Here, we adopt the
interpretation of $\ptnl$ as a potential field so that for a fixed
$\ptnl$, the time-derivative of the potential energy induced by a
velocity field, $v$, is equal to minus the power expended by the
field. We obtain the form
\begin{equation}
P_{\body}=\fc_{\body}(v)=-\sum_{0\les k\les r}\int_{\body}\gf_{,i_{1}\dots i_{k}i_{k+1}}v^{i_{k+1}}\dee{\pole^{i_{1}\dots i_{k}}}.\label{eq:power-1}
\end{equation}
It is noted that by Equation (\ref{eq:power-1}), the velocity field
enables a $k$-th order component of a multipole to sample the $(k+1)$-partial
derivatives of the potential as if it were a $(k+1)$-th component
of an $(r+1)$-multipolar distribution. For example, a moving charge
distribution samples the derivatives of the potential function as
if it were a dipole distribution. Since the power is linear in $v$,
it is the result of applying a force functional, $\fc_{\body}$, to
the velocity field. Thus, one can compute the forces and (hyper-)
stresses that the field exerts on the multipolar medium.

A related subject, considered in Sections \ref{sec:Standard-Balance}
and \ref{sec:Higher-Order-Fluxes}, is that of (hyper-) fluxes of
order $r$. Classically, the balance of an extensive property is expressed
by the balance differential equation for its flux vector field, $u$,
as $\avf_{,i}^{i}+\vform=\pform$ and boundary condition $\sform=\avf\cdot\nor$,
where $\vform$ is the time rate of the density, $\pform$ is the
source, $\sform$ is the flux density on the boundary, and $\nor$
is the unit normal to the boundary. These may be replaced by the variational
form
\begin{equation}
P_{\body}=\int_{\bdry\body}\sform\ptnl\dee A+\int_{\body}\vform\ptnl\dee V=\int_{\body}\pform\ptnl\dee V+\int_{\body}\avf^{i}\ptnl_{,i}\dee V,\label{eq:variat_Bal-2}
\end{equation}
for any test function, $\gf$, which may be interpreted as a potential
function. Hence, $P_{\body}$ may be interpreted as the power associated
with the transfer of the property. The last equation may be generalized
to 
\begin{equation}
P_{\body}=\Hfl_{\body}(\gf)=\sum_{0\les k\les r}\int_{\body}\gf_{,i_{1}\dots i_{k}}\dee{\ms s^{i_{1}\dots i_{k}}},
\end{equation}
for a given collection of Borel measures $\ms s^{i_{1}\dots i_{k}}$,
$0\les k\les r$. The functional $\Phi$ is referred to as a hyperflux
of order $r$. In the smooth case, the measures $\ms s^{i_{1}\dots i_{k}}$
are represented by smooth densities, $s^{i_{1}\dots i_{k}}$, relative
to the volume measure. For example, we show that a moving polarized
medium induces fluxes per unit length of the edges of a region.

Finally, in Section \ref{sec:Geometric-Generalizations}, we discuss
the generalization of all the above to the case where the physical
space is taken as a general differentiable manifold. While the basic
definitions may be extended to manifolds using the notion of the jet
of a function, in general, one cannot isolate the components of a
given order $k$, invariantly, that is the collections, $\{\gf_{,i_{1}\dots i_{k}}\}$
or $\{\ms s^{i_{1}\dots i_{k}}\}$.

\section{Multipole Distribution}

Let the physical space be modeled geometrically by $\rthree$. Our
basic object is \emph{an $r$-pole distribution} (often referred to
as a $2^{r}$-pole distribution), $\Charge$. Let $\testf(\rthree)$
be the space of test functions on $\rthree$, that is, the space of
compactly supported infinitely smooth functions. One may interpret
a test function, $\gf\in\testf(\rthree)$ as a variation of the potential
function for some extensive property, $\expr$. Alternatively, the
function $\gf$ may be interpreted as a potential distribution (rather
than a variation thereof) with the convention that the potential at
infinity is set to zero. A natural example is the electric potential
for the electric charge extensive property. However, the interpretation
of $\ptnl$ as a potential function for a conservative system is not
needed up to Section \ref{sec:Forces}.
\begin{defn}
\label{def:multipoes}An $r$-pole distribution is a bounded linear
functional
\begin{equation}
Q:\testf(\rthree)\tto\reals
\end{equation}
that is given in terms of a given collection of Borel measures $q,\,q^{i_{1}},\,q^{i_{1}i_{2}},\,\dots,\,q^{i_{1}\dots i_{r}}$,
$i_{k}=1,2,3$, in the form
\begin{equation}
\Charge(\gf)=\sum_{0\les k\les r}\int_{\rthree}\gf_{,i_{1}\dots i_{k}}\dee{\pole^{i_{1}\dots i_{k}}}.\label{eq:definition}
\end{equation}
If $\ptnl$ is interpreted as a (variation of the) potential function,
the action $U=\Charge(\gf)$ is interpreted as the (variation of the)
potential energy. For a given $\Charge$, the collection of measures
$\{q^{i_{1}\dots i_{k}}\}$ will be referred to the the $k$-th \emph{order
component} of the multipole distribution.
\end{defn}

It follows from the definition that a multipolar distribution is a
particular distribution in the sense of Schwartz \cite{Schwartz1966-short}.
In the terminology of \cite{Segev_Book_2023}, the multipolar distribution
is a generalized force, where a generalized velocity is taken as a
(potential) function. The measures $\pole^{i_{1}\dots i_{k}}$ are
analogous to stress measures representing the generalized force. The
representation of a multipole distribution by measures enables one
to consider singular distributions, such as a point charge or a single
dipole, in one setting with continuous distributions of charges or
multipoles. For a smooth distribution of multipoles of order $r$,
$\Charge$ is represented by smooth functions $\pdens^{i_{1}\dots i_{k}}$
such that
\begin{equation}
\Charge(\gf)=\sum_{0\les k\les r}\int_{\rthree}\gf_{,i_{1}\dots i_{k}}\pdens^{i_{1}\dots i_{k}}\dee V.
\end{equation}
In addition, the fact that the functional is represented by some given
measures, makes it possible to restrict the distribution $\Charge$
to a distribution $\Charge_{\body}$, for any Borel measurable subset
$\body$, by
\begin{equation}
\Charge_{\body}(\gf):=\sum_{0\les k\les r}\int_{\body}\gf_{,i_{1}\dots i_{k}}\dee{\pole^{i_{1}\dots i_{k}}}\label{eq:Q_b}
\end{equation}
and for the case of smooth distributions,
\begin{equation}
\Charge_{\body}(\gf)=\sum_{0\les k\les r}\int_{\body}\gf_{,i_{1}\dots i_{k}}\pdens^{i_{1}\dots i_{k}}\dee V.\label{eq:def-smooth}
\end{equation}
Henceforth, in the case of smooth multipole distributions, we will
consider regions $\body$ that are smooth, three-dimensional, compact
submanifolds of $\rthree$ with Lipschitz boundaries so that the Green-Gauss
theorem is applicable.

The zeroth component of $\Charge$, denoted by $\pole$, is interpreted
as the free charge distribution. The measures $\pole^{i}$ represent
the distribution of polarization.
\begin{example}
As an example, for a smooth distribution of dipoles,
\begin{equation}
Q_{\body}(\gf)=\int_{\body}\gf_{,i}\pdens^{i}\dee V.
\end{equation}
Using the Gauss theorem and denoting the boundary unit normal by $\nor$,
\begin{equation}
\begin{split}Q_{\body}(\gf) & =\int_{\body}(\gf\pdens^{i})_{,i}\dee{V-\int_{\body}\gf\pdens_{,i}^{i}\dee{V,}}\\
 & =\int_{\bdry\body}\gf\pdens^{i}\nor_{i}\dee{A-\int_{\body}\gf\pdens_{,i}^{i}\dee{V,}}
\end{split}
\end{equation}
so that $\pdens^{i}\nor_{i}$ is the bound charge on the boundary
and $-\pdens_{,i}^{i}$ is the distribution of the bound charge in
$\body$.
\end{example}

\section{The Bound Charge and Bound Dipole of a Quadrupole Density\label{sec:Bound-Charge-qudrupoles}}

By definition, for the case of a smooth distribution $\pdens^{ij}$
of quadruples,
\begin{equation}
Q_{\body}(\gf)=\int_{\body}\gf_{,ij}\pdens^{ij}\dee V.\label{eq:Quadrupole}
\end{equation}
Thus, we may write
\begin{equation}
\begin{split}Q_{\body}(\gf) & =\int_{\body}(\ptnl_{,i}\pdens^{ij})_{,j}\dee V-\int_{\body}\pdens_{,j}^{ij}\ptnl_{,i}\dee V,\\
 & =\int_{\bdry\body}\pdens^{ij}\nor_{j}\ptnl_{,i}\dee A-\int_{\body}\pdens_{,j}^{ij}\ptnl_{,i}\dee V.
\end{split}
\label{eq:P-2-1-1}
\end{equation}
Here, the term $-\pdens_{,j}^{ij}$ may be viewed as a bound dipole
density. The term
\begin{equation}
\pdens_{\bdry}^{i}:=\pdens^{ij}\nor_{j}\label{eq:rho_dot_n}
\end{equation}
is viewed as the components of a surface distribution, $\pdens_{\bdry}$,
of dipole density per unit area.

One may decompose $\pdens_{\bdry}$ and $\nabla\gf$ into normal and
tangential components in the form
\begin{equation}
\pdens_{\bdry}=\pdens_{\bdry\nor}+\pdens_{\bdry t},\qquad\nabla\ptnl=\nabla_{\nor}\ptnl+\nabla_{t}\ptnl\label{eq:decomp_n_t-1}
\end{equation}
 so that 
\begin{equation}
\pdens_{\bdry}\cdot\nabla\ptnl=\pdens_{\bdry\nor}\cdot\nabla_{\nor}\ptnl+\pdens_{\bdry t}\cdot\nabla_{t}\ptnl.
\end{equation}
We conclude that the first integral in the second line above may be
rewritten as
\begin{equation}
\begin{split}\int_{\bdry\body}\pdens^{ij}\nor_{j}\ptnl_{,i}\dee A & =\int_{\bdry\body}\pdens_{\bdry t}\cdot\nabla_{t}\ptnl\dee A+\int_{\bdry\body}\pdens_{\bdry\nor}\nabla_{\nor}\ptnl\dee A.\end{split}
\label{eq:decomp_integral-1}
\end{equation}

We now use the surface divergence, $\nabla_{t}\cdot v_{t}$ for a
tangent vector field $v_{t}$ on $\bdry\body$, and the surface Gauss
theorem on the boundary to rewrite the last equation as
\begin{equation}
\begin{split}\int_{\bdry\body}\pdens^{ij}\nor_{j}\ptnl_{,i}\dee A & =\int_{\bdry\body}\nabla_{t}\cdot(\pdens_{\bdry t}\ptnl)\dee{A-\int_{\bdry\body}(\nabla_{t}\cdot\pdens_{\bdry t})\ptnl\dee A}+\int_{\bdry\body}\pdens_{\bdry\nor}\nabla_{\nor}\ptnl\dee A.\end{split}
\label{eq:surface_term}
\end{equation}
In case $\body$ has a smooth boundary, $\bdry(\bdry\body)=\varnothing$,
so, by the divergence theorem for surfaces, the first integral vanishes,
and 
\begin{equation}
\int_{\bdry\body}\pdens^{ij}\nor_{j}\ptnl_{,i}\dee A=-\int_{\bdry\body}(\nabla_{t}\cdot\pdens_{\bdry t})\ptnl\dee A+\int_{\bdry\body}\pdens_{\bdry\nor}\nabla_{\nor}\ptnl\dee A.
\end{equation}

Integrating by parts the second integral in (\ref{eq:P-2-1-1}),
\begin{equation}
\int_{\body}\pdens_{,j}^{ij}\ptnl_{,i}\dee V=\int_{\bdry\body}\pdens_{,j}^{ij}\nor_{i}\ptnl\dee A-\int_{\body}\pdens_{,ij}^{ij}\ptnl\dee V,
\end{equation}
and so, for the case where $\bdry\body$ is smooth, 
\begin{equation}
Q_{\body}(\gf)=-\int_{\bdry\body}(\nabla_{t}\cdot\pdens_{\bdry t})\ptnl\dee A+\int_{\bdry\body}\pdens_{\bdry\nor}\nabla_{\nor}\ptnl\dee{A+\int_{\body}\pdens_{,ij}^{ij}\ptnl\dee V-\int_{\bdry\body}\pdens_{,j}^{ij}\nor_{i}\ptnl\dee A.}\label{eq:quad-smooth}
\end{equation}
The term $-\textrm{div}_{t}\pdens_{\bdry t}$ is interpreted as the
bound charge density per unit area on the surface.

When the boundary $\bdry\body$ is not continuously differentiable,
the unit normal is not continuous. In particular, the unit normal
has a jump discontinuity along an edge. Consequently, $\pdens^{ij}\nor_{j}$
has a jump discontinuity along an edge, resulting in a concentration
of density of the property per unit length along the edge. To get
some insight to the situation where $\bdry\body$ is not continuously
differentiable, we consider the case where $\body$ is an oriented
simplex so that (see \cite{Segev_Book_2023} for the notation)
\begin{equation}
\body=[x_{0},x_{1},x_{2},x_{3}],\qquad x_{a}\in\rthree,\;a=0,1,2,3.
\end{equation}
The $a$-th face of $\body$, $a=0,\dots,3$ is the two dimensional
simplex 
\begin{equation}
\bdry_{a}:=(-1)^{a}[x_{0},\dots,\hat{x}_{a},\dots,x_{3}],
\end{equation}
where a superimposed ``hat'' indicates the omission of a term. As
a chain, 
\begin{equation}
\bdry\body=\sum_{a=0}^{3}\bdry\body_{a}.
\end{equation}

Inductively, the $b$-th edge of $\bdry\body_{a}$, $\bdry\body_{ab}$,
$b=0,\dots,3$, $b\ne a$, is given by
\begin{equation}
\bdry\body_{ab}=\begin{cases}
(-1)^{a+b}[x_{0},\dots,\hat{x}_{b},\dots,\hat{x}_{a},\dots,x_{3}], & \text{if }b<a,\\
(-1)^{a+b+1}[x_{0},\dots,\hat{x}_{a},\dots,\hat{x}_{b},\dots,x_{3}], & \text{if }b>a.
\end{cases}
\end{equation}
 Evidently, $\bdry\body_{ab}$ contains the same points as $\bdry\body_{ba}$
and has the opposite orientation, that is, 
\begin{equation}
\bdry\body_{ab}=-\bdry\body_{ba}.
\end{equation}

For the case of a simplex, we denote the normal to $\bdry\body_{a}$
by $\nor^{a}$, and we rewrite Equation (\ref{eq:rho_dot_n}), etc.,
as follows. Set
\begin{equation}
\pdens_{a}^{i}:=\pdens^{ij}\nor_{j}^{a},\qquad\pdens_{a}=\pdens_{a\nor}+\pdens_{at},
\end{equation}
so that
\begin{equation}
\pdens_{a}\cdot\nabla\ptnl=\pdens_{a\nor}\nabla_{\nor}\ptnl+\pdens_{at}\cdot\nabla_{t}\ptnl.
\end{equation}
Thus, Equation (\ref{eq:decomp_integral-1}) is rewritten as 
\begin{equation}
\int_{\bdry\body}\pdens^{ij}\nor_{j}\ptnl_{,i}\dee A=\sum_{a=0}^{3}\int_{\bdry\body_{a}}\pdens_{at}\cdot\nabla_{t}\ptnl\dee A+\sum_{a=0}^{3}\int_{\bdry\body_{a}}\pdens_{a\nor}\nabla_{\nor}\ptnl\dee A.
\end{equation}
Hence,
\begin{equation}
\begin{split}\int_{\bdry\body}\pdens^{ij}\nor_{j}\ptnl_{,i}\dee A & =\sum_{a=0}^{3}\int_{\bdry\body_{a}}\nabla_{t}\cdot(\ptnl\resto{\bdry\gO_{a}}\pdens_{at})\dee{A^{a}}\\
 & \qquad\qquad-\sum_{a=0}^{3}\int_{\bdry\body_{a}}(\nabla_{t}\cdot\pdens_{at})\ptnl\resto{\bdry\gO_{a}}\dee{A^{a}}\\
 & \qquad\qquad\qquad\qquad+\sum_{a=0}^{3}\int_{\bdry\body_{a}}\pdens_{a\nor}\nabla_{\nor}\ptnl\dee A,
\end{split}
\end{equation}
Noting that the sum of integrals on the first line above does not
vanish identically, we let $\mu_{ab}$ be the outwards pointing unit
normal to $\bdry\body_{ab}$ in $\bdry\body_{a}$. Using the surface
divergence theorem, one has
\begin{equation}
\begin{split}\sum_{a=0}^{3}\int_{\bdry\body_{a}}\nabla_{t}\cdot(\ptnl\resto{\bdry\gO_{a}}\pdens_{at})\dee{A^{a}} & =\sum_{\underset{a\ne b}{a,b=0}}^{3}\int_{\bdry\body_{ab}}\pdens_{at}\cdot\mu_{ab}\ptnl\dee{L^{ab}},\\
 & =\sum_{a>b}\int_{\bdry\body_{ab}}(\pdens_{at}\cdot\mu_{ab}-\pdens_{bt}\cdot\mu_{ba})\ptnl\dee{L^{ab}}.
\end{split}
\end{equation}

Since the expression $\pdens_{at}\cdot\mu_{ab}-\pdens_{bt}\cdot\mu_{ba}$
multiplies the function $\ptnl$, 
\begin{equation}
l_{ab}:=\pdens_{at}\cdot\mu_{ab}-\pdens_{bt}\cdot\mu_{ba},\qquad a>b,\label{eq:l_ab-1}
\end{equation}
is interpreted as the density of the property per unit length along
the edge $\bdry\body_{ab}$. It is noted that we can also write 
\begin{equation}
l_{ab}:=\pdens_{a}\cdot\mu_{ab}-\pdens_{b}\cdot\mu_{ba},\qquad a>b,\label{eq:l_ab-2}
\end{equation}
because $\pdens_{a\nor}\cdot\mu_{ab}=\pdens_{b\nor}\mu_{ba}=0$.

We conclude that the expression for the power in Equation (\ref{eq:quad-smooth})
is accordingly extended and assumes the form
\begin{multline}
Q_{\body}(\gf)=-\int_{\bdry\body}(\nabla_{t}\cdot\pdens_{\bdry t})\ptnl\dee A+\int_{\bdry\body}\pdens_{\bdry\nor}\nabla_{\nor}\ptnl\dee{A+\int_{\body}\pdens_{,ij}^{ij}\ptnl\dee V}\\
-\int_{\bdry\body}\pdens_{,j}^{ij}\nor_{i}\ptnl\dee A+\sum_{a>b}\int_{\bdry\body_{ab}}l_{ab}\ptnl\dee{L^{ab}}.\label{eq:quad-simplex}
\end{multline}

\section{Forces on Multi-Polarized Bodies\label{sec:Forces}}

We consider multipole distributions as above. So far, the interpretation
of $\gf$ as a variation of the potential function was not necessary.
Moreover, no physical consequence has been deduced from such an interpretation.
Now, we adopt the interpretation of $\ptnl$ as a potential, and we
postulate that the mechanical work $W_{\body}$, performed by the
mechanical force under a motion of the multipolar distribution distribution
carried with the motion of the material in $\body$, is equal to minus
the change in the potential energy $\Charge_{\body}(\ptnl)$(\ref{eq:Q_b}).
It is emphasized that we take the notion of mechanical power as fundamental
and a mechanical force is viewed as a bounded linear functional defined
on the topological vector space of velocity fields defined on the
body (see \cite{Segev_Book_2023} for details and applications of
this framework). It is also noticed that the mechanical force acts
on the velocity field of the material body in which the multipolar
distribution is embedded.

In this section, we assume that a body is a compact three-dimensional
submanifold with corners of $\rthree$. A transport of multipoles
in a region, $\body$, in space is assumed to be specified by an embedding
\begin{equation}
\conf:\body\tto\rthree.
\end{equation}
A motion of the multipoles in the region $\body$ is a smooth mapping
\begin{equation}
c:\reals\times\body\tto\rthree,
\end{equation}
such that 
\begin{equation}
c_{t}:=c\resto{\{t\}\times\body}:\body\tto\rthree
\end{equation}
is an embedding. For convenience, we set $c_{0}(x)=x$ for all $x\in\body$.

At any time $t$, the multipolar distribution $\Charge_{\body}$ induces
a multipolar distribution $c_{t*}(\Charge_{\body})$ defined on $c_{t}(\body)$,
the \emph{pushforward of the distribution} by $c_{t}$, defined by
\begin{equation}
c_{t*}(\Charge_{\body})(\ptnl):=\Charge_{\body}(\ptnl\comp c_{t}).
\end{equation}
The pushforward, $c_{t*}(\Charge_{\body})$, represents the image
of the multipolar distribution as it is carried with the deformation
$c_{t}$. This is a special case of the operation of a pushforward
of a de Rham current (a zero current in our case) by a mapping (see
\cite[p.~47]{deRham1955}). If $c_{t*}(\Charge_{\body})$ is represented
by the measures $p^{i_{1}\dots i_{k}}$, then the potential energy
of the material body at time $t$ is
\begin{equation}
c_{t*}(\Charge_{\body})(\ptnl)=\sum_{0\les k\les r}\int_{c(\body)}\gf_{,i_{1}\dots i_{k}}\dee{p^{i_{1}\dots i_{k}}}=\sum_{0\les k\les r}\int_{\body}(\gf\comp c_{t})_{,i_{1}\dots i_{k}}\dee{\pole^{i_{1}\dots i_{k}}}.
\end{equation}
The specific expression for the measures $p^{i_{1}\dots i_{k}}$ is
somewhat lengthy as it requires a repeated use of the chain rule.
The repeated use of the chain rule also implies that in general, $p^{i_{1}\dots i_{k}}$
depends on $\pole^{i_{1}\dots i_{l}}$ for $l>k$. In other words,
the $k$-th order components are not invariant under a deformation.
See an analogous remark in Section \ref{sec:Geometric-Generalizations}.

It follows that the change in the potential energy of the material
body between time zero and time $t$ is given by 

\begin{equation}
\begin{split}\Delta U_{t} & =Q_{\body}(\ptnl\comp c_{t})-\Charge_{\body}(\ptnl),\\
 & =\sum_{0\les k\les r}\int_{\body}(\gf\comp c_{t})_{,i_{1}\dots i_{k}}\dee{\pole^{i_{1}\dots i_{k}}}-\sum_{0\les k\les r}\int_{\body}\gf_{,i_{1}\dots i_{k}}\dee{\pole^{i_{1}\dots i_{k}},}\\
 & =\sum_{0\les k\les r}\int_{\body}(\gf\comp c_{t}-\gf)_{,i_{1}\dots i_{k}}\dee{\pole^{i_{1}\dots i_{k}}}.
\end{split}
\end{equation}

Since the change of potential energy is equal to minus the mechanical
work, the mechanical power at time zero is given by
\begin{equation}
P_{\body}=-\frac{d}{\dee t}\left[\sum_{0\les k\les r}\int_{\body}(\gf\comp c_{t}-\gf)_{,i_{1}\dots i_{k}}\dee{\pole^{i_{1}\dots i_{k}}}\right]_{t=0}.
\end{equation}
Using the chain-rule and writing $\gf(x^{j}(t))$ for $\gf\comp c_{T}$
\begin{equation}
\begin{split}P_{\body} & =-\frac{d}{\dee t}\left[\sum_{0\les k\les r}\int_{\body}(\gf(x^{j}(t)))_{,i_{1}\dots i_{k}}\dee{\pole^{i_{1}\dots i_{k}}}\right]_{t=0},\\
 & =-\left[\sum_{0\les k\les r}\int_{\body}\frac{d}{\dee t}(\gf_{,i_{1}\dots i_{k}}(x^{j}(t)))\dee{\pole^{i_{1}\dots i_{k}}}\right]_{t=0},\\
 & =-\left[\sum_{0\les k\les r}\int_{\body}\gf_{,i_{1}\dots i_{k}i_{k+1}}(x^{j}(t))\frac{dx^{i_{k+1}}}{\dee t}(t)\dee{\pole^{i_{1}\dots i_{k}}}\right]_{t=0}.
\end{split}
\end{equation}
We conclude that
\begin{equation}
P_{\body}=-\sum_{0\les k\les r}\int_{\body}\gf_{,i_{1}\dots i_{k}i_{k+1}}v^{i_{k+1}}\dee{\pole^{i_{1}\dots i_{k}}},\label{eq:power}
\end{equation}
where, 
\begin{equation}
v^{i_{k+1}}(x)=\frac{dx^{i_{k+1}}}{\dee t}(t=0)
\end{equation}
is the corresponding velocity field.

It is noted that by Equation (\ref{eq:power}) the velocity field
enables a $k$-th order component of a multipole to sample the $(k+1)$-partial
derivatives of the potential, as if it were a $(k+1)$-th component.
For example, a moving charge distribution samples the derivatives
of the potential function as if it were a dipole distribution.

The expression above represents an action of a linear functional $F_{\body}$
defined on velocity fields $v$; hence, it is interpreted as a generalized
force. We write therefore,
\begin{equation}
P_{\body}=\fc_{\body}(v)=-\int_{\body}v^{j}\sum_{0\les k\les r}\gf_{,ji_{1}\dots i_{k}}\dee{\pole^{i_{1}\dots i_{k}}}.\label{eq:force}
\end{equation}

For the case where the measures $q^{i_{1}\dots i_{k}}$ are given
in terms of smooth densities $\pdens^{i_{1}\dots i_{k}}$,
\begin{equation}
\begin{split}\fc_{\body}(v) & =-\int_{\body}v^{j}\sum_{0\les k\les r}\gf_{,ji_{1}\dots i_{k}}\pdens^{i_{1}\dots i_{k}}\dee V,\\
 & =-\int_{\bdry\body}v^{j}\sum_{0\les k\les r}\gf_{,ji_{1}\dots i_{k-1}}\pdens^{i_{1}\dots i_{k}}\nor_{i_{k}}\dee A+\int_{\body}v^{j}\sum_{0\les k\les r}\gf_{,ji_{1}\dots i_{k-1}}\pdens_{,i_{k}}^{i_{1}\dots i_{k}}\dee V\\
 & \qquad\qquad\qquad+\int_{\body}v_{,i_{k}}^{j}\sum_{0\les k\les r}\gf_{,ji_{1}\dots i_{k-1}}\pdens^{i_{1}\dots i_{k}}\dee V.
\end{split}
\end{equation}
Note that the term multiplying $v_{,i_{k}}^{j}$ represents a stress
tensor.
\begin{example}
\label{exa:Force_on_Polarized}For a body $\body\subset\rthree$,
which is smoothly polarized by density $\pdens^{i}$, the variation
potential energy is given by
\begin{equation}
U=\int_{\body}\gf_{,i}\pdens^{i}\dee V,
\end{equation}
and for a velocity field $v:\body\tto\rthree,$
\begin{equation}
\begin{split}\dot{U}=-P & =\frac{d}{\dee t}\int_{\body}\left[\gf(x^{j}+tv^{j})_{,i}\pdens^{i}\right]\dee V,\\
 & =\int_{\body}\gf_{,ij}v^{j}\pdens^{i}\dee V,\\
 & =\int_{\body}\gf_{,ji}v^{j}\pdens^{i}\dee V,\\
 & =\int_{\bdry\body}\gf_{,j}\pdens^{i}\nor_{i}v^{j}\dee A-\int_{\body}\gf_{,j}\pdens_{,i}^{i}v^{j}\dee V-\int_{\body}\gf_{,j}\pdens^{i}v_{,i}^{j}\dee V.
\end{split}
\end{equation}
We view the first integral as minus the power of the force exerted
by the electric field on the bound charge induced by the dielectric
on the boundary, the second integral is the power of the force exerted
on the induced bound charge density in the dielectric, and the third
integral as the stress $\gf_{,j}\pdens^{i}$ acting on the velocity
gradient. It is observed that the stress tensor is not symmetric and
its skew symmetric component, representing a couple density, acts
on the angular velocity component of the velocity gradient.
\end{example}

\begin{example}
We now consider the forces acting on a body in which a quadrupole
is smoothly distributed with density $\pdens^{ij}$ per unit volume.
In view of (\ref{eq:force}),
\begin{equation}
\fc_{\body}(v)=\int_{\body}v^{i}\gf_{,ijk}\pdens^{jk}\dee V.
\end{equation}

An integration by parts procedure analogous to that of Section \ref{sec:Bound-Charge-qudrupoles}
gives
\begin{equation}
\fc_{\body}(v)=\int_{\bdry\body}v^{i}\gf_{,ij}\pdens^{jk}\nor_{k}\dee A-\int_{\body}v_{,k}^{i}\gf_{,ij}\pdens^{jk}\dee V-\int_{\body}v^{i}\gf_{,ij}\pdens_{,k}^{jk}\dee V.\label{eq:111}
\end{equation}
For the first integral, we have for the case of a simplex, 
\begin{multline}
\int_{\bdry\body}v^{i}\gf_{,ij}\pdens^{jk}\nor_{k}\dee A\\
\begin{split} & =\int_{\bdry\body}(v^{i}\gf_{,i})_{,j}\pdens_{\bdry}^{j}\dee A-\int_{\bdry\body}v_{,j}^{i}\gf_{,i}\pdens_{\bdry}^{j}\dee A,\\
 & =\int_{\bdry\body}\nabla_{t}(v^{i}\gf_{,i})\cdot\pdens_{\bdry t}\dee A+\int_{\bdry\body}\nabla_{\nor}(v^{i}\gf_{,i})\cdot\pdens_{\bdry\nor}\dee A-\int_{\bdry\body}v_{,j}^{i}\gf_{,i}\pdens_{\bdry}^{j}\dee A,\\
 & =\int_{\bdry\body}\nabla_{t}\cdot(\pdens_{\bdry t}v^{i}\ptnl_{,i})\dee A-\int_{\bdry\body}(\nabla_{t}\cdot\pdens_{\bdry t})v^{i}\ptnl_{,i}\dee A\\
 & \qquad\qquad+\int_{\bdry\body}\nabla_{\nor}(v^{i}\gf_{,i})\cdot\pdens_{\bdry\nor}\dee A-\int_{\bdry\body}v_{,j}^{i}\gf_{,i}\pdens_{\bdry}^{j}\dee A\\
 & =\sum_{a=0}^{3}\int_{\bdry\body_{a}}\nabla_{t}\cdot(\pdens_{at}v^{i}\ptnl_{,i})\dee{A^{a}}-\int_{\bdry\body}(\nabla_{t}\cdot\pdens_{\bdry t})v^{i}\ptnl_{,i}\dee A\\
 & \qquad\qquad+\int_{\bdry\body}\nabla_{\nor}(v^{i}\gf_{,i})\cdot\pdens_{\bdry\nor}\dee A-\int_{\bdry\body}v_{,j}^{i}\gf_{,i}\pdens_{\bdry}^{j}\dee A.
\end{split}
\end{multline}
By the surface divergence theorem, and using Cartesian coordinates
on the faces,
\begin{equation}
\begin{split}\sum_{a=0}^{3}\int_{\bdry\body_{a}}\nabla_{t}\cdot(\pdens_{at}v^{i}\ptnl_{,i})\dee{A^{a}} & =\sum_{\underset{a\ne b}{a,b=0}}^{3}\int_{\bdry\body_{ab}}v^{i}\ptnl_{,i}\pdens_{at}\cdot\mu_{ab}\dee{L^{ab}},\\
 & =\sum_{a>b}\int_{\bdry\body_{ab}}v^{i}\ptnl_{,i}(\pdens_{at}\cdot\mu_{ab}-\pdens_{bt}\cdot\mu_{ba})\dee{L^{ab}}.
\end{split}
\end{equation}
Since they multiply the velocity field $v$, the expressions
\begin{equation}
F_{i}^{L}:=\ptnl_{,i}(\pdens_{at}\cdot\mu_{ab}-\pdens_{bt}\cdot\mu_{ba})
\end{equation}
are viewed as the components of a force density $F^{L}$ per unit
length concentrated along the edges.

In addition, for the second integral in (\ref{eq:111}), we write
\begin{equation}
\begin{split}\int_{\body}v_{,k}^{i}\gf_{,ij}\pdens^{jk}\dee V & =\int_{\bdry\body}v_{,k}^{i}\gf_{,i}\pdens^{jk}\nor_{j}\dee A-\int_{\body}v_{,jk}^{i}\gf_{,i}\pdens^{jk}\dee V-\int_{\body}v_{,k}^{i}\gf_{,i}\pdens_{,j}^{jk}\dee V\end{split}
,
\end{equation}
and for the third integral
\begin{equation}
\int_{\body}v^{i}\gf_{,ij}\pdens_{,k}^{jk}\dee V=\int_{\bdry\body}v^{i}\gf_{,i}\pdens_{,k}^{jk}\nor_{j}\dee A-\int_{\body}v_{,j}^{i}\gf_{,i}\pdens_{,k}^{jk}\dee V-\int_{\body}v^{i}\gf_{,i}\pdens_{,jk}^{jk}\dee V.
\end{equation}

We conclude that
\begin{multline}
\fc_{\body}(v)\\
\begin{split} & =\sum_{a>b}\int_{\bdry\body_{ab}}v^{i}F_{i}^{L}\dee{L^{ab}}-\int_{\bdry\body}(\nabla_{t}\cdot\pdens_{\bdry t})v^{i}\ptnl_{,i}\dee A+\int_{\bdry\body}\nabla_{\nor}(v^{i}\gf_{,i})\cdot\pdens_{\bdry\nor}\dee A\\
 & \qquad-\int_{\bdry\body}v_{,j}^{i}\gf_{,i}\pdens_{\bdry}^{j}\dee A-\int_{\bdry\body}v_{,k}^{i}\gf_{,i}\pdens^{jk}\nor_{j}\dee A+\int_{\body}v_{,jk}^{i}\gf_{,i}\pdens^{jk}\dee V\\
 & \qquad+\int_{\body}v_{,k}^{i}\gf_{,i}\pdens_{,j}^{jk}\dee V-\int_{\bdry\body}v^{i}\gf_{,i}\pdens_{,k}^{jk}\nor_{j}\dee A+\int_{\body}v_{,j}^{i}\gf_{,i}\pdens_{,k}^{jk}\dee V
\end{split}
\\
+\int_{\body}v^{i}\gf_{,i}\pdens_{,jk}^{jk}\dee V.
\end{multline}
It is observed that the terms involving $v_{,j}^{i}$ indicate stresses
and the terms involving $v_{,jk}^{i}$ indicate hyper-stresses.
\end{example}

\section{Standard Balance of Scalar Properties\label{sec:Standard-Balance}}

Let $\expr$ be a smoothly distributed extensive property transported
in space, $\rthree$. Let $\dens$ be the volume density distribution,
$\vform:=\bdry\dens/\bdry t$, and let $\pform$ be the source density.
The flux of the property through the boundary is described by a field
$\sform$ over $\bdry\gO$. Let $\body\subset\rthree$ be a generic
regular bounded domain, say a bounded open set with a Lipschitz boundary
(i.e., the boundary is a Lipschitz surface). The balance of the property
in the region $\body$ is expressed by the equation
\begin{equation}
\int_{\bdry\body}\sform\dee A+\int_{\body}\vform\dee V=\int_{\body}\pform\dee V.\label{eq:Int_Bal-1}
\end{equation}

Assuming that the Cauchy postulates hold, there is a flux vector field
$u$ defined in space, such that the flux distribution, $\tau$, on
the boundary, $\bdom$, is given by 
\begin{equation}
\sform=\avf\cdot\nor=u^{i}\nor_{i}.\label{eq:BC-1}
\end{equation}
Substituting (\ref{eq:BC-1}) into (\ref{eq:Int_Bal-1}), using the
Gauss theorem, the arbitrariness of $\body$ implies that any point
is a Lebesgue point for the integrand, i.e. 
\begin{equation}
\avf_{,i}^{i}+\vform=\pform,\label{eq:diff_Bal-1}
\end{equation}
Conversely, Equations (\ref{eq:diff_Bal-1}) and (\ref{eq:BC-1})
together, imply balance, as represented by Equation (\ref{eq:Int_Bal-1}).

Multiplying (\ref{eq:diff_Bal-1}) by a smooth function $\ptnl$ with
compact support in $\rthree$, integrating by parts, and using the
Gauss theorem, one obtains
\begin{equation}
P=\int_{\bdry\body}\sform\ptnl\dee A+\int_{\body}\vform\ptnl\dee V=\int_{\body}\pform\ptnl\dee V+\int_{\body}\avf^{i}\ptnl_{,i}\dee V.\label{eq:variat_Bal-1}
\end{equation}

A function $\ptnl$, as above, may be interpreted as a potential function.
However, we do not assume a-priori that we consider a conservative
field, and the functions play only a formal mathematical role. Evidently,
(\ref{eq:Int_Bal-1}) can be recovered from (\ref{eq:variat_Bal-1})
by setting $\ptnl=1$ in $\body$.

When $\ptnl$ is interpreted as a potential function, (\ref{eq:variat_Bal-1})
is interpreted as equality of expressions for power. The left-hand
side is interpreted as the power due to the rate of change of density
of the property and the rate at which the property exits or enters
the region. The right-hand side is interpreted as the power expended
due to sources in $\body$, a source of energy, plus the power due
to the flow of the property along the gradient of the ``potential''
field. It is observed that unlike the left expression for $P$, the
expression on the right can be naturally restricted to any subregion
$\mc R\subset\body$ in the form
\begin{equation}
P_{\mc R}=\int_{\mc R}\pform\ptnl\dee V+\int_{\mc R}\avf^{i}\ptnl_{,i}\dee V.\label{eq:power_smooth_case}
\end{equation}
Moreover, the last expression may be generalized to the case of Borel
measures, $\ms{\pform}$, $\ms{\avf}^{i}$, so that 
\begin{equation}
P_{\mc R}=\int_{\mc R}\ptnl\dee{\ms{\pform}}+\int_{\mc R}\ptnl_{,i}\dee{\ms{\avf}^{i}},\label{eq:power_for_measures}
\end{equation}
thus allowing the flux distribution to be singular. This generalized
form is meaningful when $\mc R$ is any measurable set, so the normal
to the boundary $\bdry\mc R$ need not be defined.

Consequently, we now change the point of view and take $P$ as a fundamental
notion. Thus, we assume from the start that there is a scalar Borel
measure $\ms{\pform}$ and a vector-valued measure $\ms u$ such that
the power expended in the region $\mc R$ is given by (\ref{eq:power_for_measures}).
\begin{equation}
P_{\mc R}=\int_{\mc R}\ptnl\dee{\ms{\pform}}+\int_{\mc R}\ptnl_{,i}\dee{\ms{\avf}^{i}}.\label{eq:Ass_Power-1}
\end{equation}
For the smooth case, $\mc R$ is assumed to have a Lipschitz boundary
and the measures are given by differentiable densities $\pform$ and
$u^{i}$ as in Equation (\ref{eq:power_smooth_case}). In this case,
tracing back the derivation of (\ref{eq:variat_Bal-1}), one can retrieve
the equality of the expressions for the power as in Equation (\ref{eq:variat_Bal-1})
where $\vform$ and $\sform$ are defined by (\ref{eq:diff_Bal-1})
and (\ref{eq:BC-1}), respectively.

\section{Higher-Order Fluxes\label{sec:Higher-Order-Fluxes}}

The expression for the power in Equation (\ref{eq:Ass_Power-1}) suggests
a natural extension to hyperfluxes of order $r$.
\begin{defn}
\label{def:hyperflux}A \emph{flux (hyperflux) of order $r$} \emph{of
an extensive property $\expr$ }is a linear functional $\Hfl$ on
the space of infinitely smooth, compactly supported, real-valued functions
given in terms of a collection of Borel Measures $\ms s^{i_{1}\dots i_{k}}$,
$0\les k\les r$, in the form
\begin{equation}
\Hfl(\ptnl)=\sum_{0\les k\les r}\int_{\rthree}\ptnl_{,i_{1}\dots i_{k}}\dee{\ms s^{i_{1}\dots i_{k}}}.\label{eq:Def_Hflux}
\end{equation}
Note that the flux functional may be naturally restricted to any Borel
measurable subset, $\body$, by
\[
\Hfl_{\body}(\ptnl)=\sum_{0\les k\les r}\int_{\body}\ptnl_{,i_{1}\dots i_{k}}\dee{\ms s^{i_{1}\dots i_{k}}}.
\]
\end{defn}

Evidently, in terms of mathematical expressions, Definition \ref{def:hyperflux}
is identical to Definition \ref{def:multipoes}, however, the physical
interpretation is different. In the definition of a multipole distribution,
a test function $\gf$ is interpreted as a variation of a potential
function for a given multipole distribution of the property $\expr$,
and the action of the multipole distribution on $\gf$ is interpreted
as power required to vary the potential. On the other hand, for the
definition of a hyperflux, the potential is assumed to be fixed and
the value $\Hfl(\ptnl)$ may be interpreted the the power expended
in the rearrangement of the property $\expr$.

In the smooth case there are smooth functions $s^{i_{1}\dots i_{k}}$
such that 
\begin{equation}
\Hfl(\ptnl)=\sum_{0\les k\les r}\int_{\rthree}\ptnl_{,i_{1}\dots i_{k}}s^{i_{1}\dots i_{k}}\dee V.\label{eq:Hflux-smooth}
\end{equation}
For example, for $k=0$, $\pform$ is the standard source term and
for $k=1$, $s^{i}$ is the smooth flux vector field denoted in the
previous section by $u^{i}$. The components $\pform^{ij}$ represent
a second-order flux (hyperflux), a tensor field that may be thought
of as a smooth field of flux dipoles. Definition \ref{def:hyperflux}
allows such fields that are singular.
\begin{rem}
When we identify the functions $\ptnl$ with potential distributions,
the total potential energy, $U$, for a distribution of the property
$\expr$, specified by a distribution of multipoles $\{q^{i_{1}\dots i_{k}}\}$,
is given as in Definition \ref{def:multipoes} by 
\begin{equation}
U=\Charge(\ptnl)=\sum_{0\les k\les r}\int_{\rthree}\ptnl_{,i_{1}\dots i_{k}}\dee{\pole^{i_{1}\dots i_{k}}}.
\end{equation}
If both the potential field, $\ptnl$, and the distribution of multipoles,
$\{q^{i_{1}\dots i_{k}}\}$, are time-dependent, differentiation of
the previous relation with respect to time gives
\begin{equation}
\dot{U}=\sum_{0\les k\les r}\int_{\rthree}\dot{\gf}_{,i_{1}\dots i_{k}}\dee{\pole^{i_{1}\dots i_{k}}}+\sum_{0\les k\les r}\int_{\rthree}\ptnl_{,i_{1}\dots i_{k}}\dee{\ms s^{i_{1}\dots i_{k}}},
\end{equation}
where a superimposed dot indicates partial differentiation relative
to the time variable and 
\begin{equation}
\ms s^{i_{1}\dots i_{k}}:=\dot{\pole}^{i_{1}\dots i_{k}}.
\end{equation}
Thus, the first sum indicates the power needed to vary the potential
energy, and the second sum indicates the power needed to rearrange
the property.
\end{rem}

\begin{example}
\label{exa:flux_of_polarized} Using Equation (\ref{eq:power}), the
power expended in the motion of a neutral polarized material with
velocity field $v$, and dipole distribution given by the measures
$q^{i}$ is given by 
\begin{equation}
P_{\body}=-\sum_{0\les k\les r}\int_{\body}\ptnl_{,ij}v^{j}\dee{\pole^{i}}.
\end{equation}
Comparison with (\ref{eq:Def_Hflux}) implies that the rearrangement
of the charge is given by
\begin{equation}
\ms s^{ij}=-v^{j}q^{i}.
\end{equation}
In the smooth case, 
\begin{equation}
P_{\body}=-\sum_{0\les k\les r}\int_{\body}\ptnl_{,ij}v^{j}\rho^{i}\dee V,
\end{equation}
which as the same form as Equation (\ref{eq:Quadrupole}) when we
make the identification 
\begin{equation}
\pform^{ij}=\rho^{ij}:=\rho^{i}v^{j}.\label{eq:quad-p-1}
\end{equation}
Thus, Equations (\ref{eq:quad-simplex},\ref{eq:l_ab-2}) hold with
\begin{equation}
\pdens_{a}^{i}:=v^{j}\rho^{i}\nor_{j}^{a}.
\end{equation}
In particular, Equation (\ref{eq:quad-simplex}) implies that charge
flows on the boundary and that a charge density per unit length $l_{ab}$
is flowing through the edge $\bdry\body_{ab}$.

\end{example}

\section{On Geometric Generalizations\label{sec:Geometric-Generalizations}}

The expression
\begin{equation}
\Charge(\gf)=\sum_{0\les k\les r}\int_{\rthree}\gf_{,i_{1}\dots i_{k}}\dee{\pole^{i_{1}\dots i_{k}}},\label{eq:form}
\end{equation}
that appears in Definition \ref{def:multipoes} for multipole distributions,
and the similar expression for hyperfluxes utilizes the geometric
structure of $\rthree$ modeling the physical space. The operator
$\Charge$ is a bounded linear functional on the space of compactly
supported $C^{r}$-functions. Hence, it is a Schwartz distribution
of order $r$.

This expression may be formally generalized to the general geometric
setting where the physical space, $\spc$, is assumed to be a general
manifold, devoid of a specific Riemannian metric or a connection.
To that end, we use the notion of a de Rham current \cite{deRham1955,Federer1969,Giaquinta1}.
It is recalled that a de Rham $p$-current of order $r$ on a manifold
$\spc$ is a linear functional defined on the topological vector space
of $C^{r}$-compactly supported differential $p$-forms. Thus, as
a test function is a $0$-form, the functional $\Charge$ is a $0$-current
of order $r$ on the manifold $\spc$.

On a general differentiable manifold, a particular partial derivative
of order $k$, $0<k\les r$, of a function is not an invariant object
(under a transformation of coordinates). Moreover, even the collection
of all partial derivatives of order $k$, $k>1$ does not represent
an invariant geometric object. (For the case $k=1$, the collection
of partial derivatives $\{\gf_{,i}\}$ represents the exterior derivative
$\dee{\gf}$.) However, the collection of all partial derivatives
of order $k$, $0<k\les r$, of a function $\gf$ represents an invariant
object\textemdash the $r$-jet extension of $\gf$, $j^{r}\gf$.\footnote{It is noted that even when a connection is given on $\spc$, the covariant
derivatives ``mix'' the partial derivatives of different orders.}

It follows that the natural generalization of (\ref{eq:form}) is
\begin{equation}
\Charge(\gf)=\int_{\spc}j^{r}\gf\cdot\dee q,\label{eq:form-manifolds}
\end{equation}
where $q$ is a measure-valued section of the dual of the jet bundle,
that is, a measure-valued section of $J^{r}(\spc,\reals)^{*}$ (see
\cite{Segev_Book_2023}). Indeed, locally, the integrand of (\ref{eq:form-manifolds})
is of the form of the integrand of (\ref{eq:form}).

It may be shown (\emph{loc. cit.}) that every $0$-current of order
$r$ on the manifold $\spc$, is of the form (\ref{eq:form-manifolds}),
for a measure-valued section of the dual of the jet bundle. However,
for a given such current, the collection of measures is not unique.
A multipole distribution, given in terms of the measures $\{\pole^{i_{1}\dots i_{k}}\}$,
specifies a particular representation of the de Rham current locally.
\begin{acknowledgement*}
The authors thank Victor Eremeyev for reading a draft of the manuscript
and making comments.
\end{acknowledgement*}
%
%\bibliographystyle{alpha}
%\bibliography{refs}

\begin{thebibliography}{TdCMG04}
	
	\bibitem[DO14]{Dorfamand_and_Ogden_2014}
	L.~Dorfmann and R.W. Ogden.
	\newblock {\em Nonlinear Theory of Electroelastic and Magnetoelastic
		Interactions}.
	\newblock Springer, 2014.
	
	\bibitem[dR84]{deRham1955}
	G.~de~Rham.
	\newblock {\em Differentiable Manifolds}.
	\newblock Springer, 1984.
	
	\bibitem[Fed69]{Federer1969}
	H.~Federer.
	\newblock {\em Geometric Measure Theory}.
	\newblock Springer, 1969.
	
	\bibitem[GMS98]{Giaquinta1}
	M.~Giaquinta, G.~Modica, and J.~Soucek.
	\newblock {\em Cartesian Currents in the Calculus of Variations {I}}.
	\newblock Springer, 1998.
	
	\bibitem[Jac99]{Jackson}
	J.D. Jackson.
	\newblock {\em Classical Electrodynamics}.
	\newblock Wiley, 1999.
	
	\bibitem[Kaf71]{Kafadar71}
	C.B. Kafadar.
	\newblock The theory of multipoles in calssical electromagnetism.
	\newblock {\em International Journal of Engineering Science}, 9:831--853, 1971.
	
	\bibitem[LL95]{LandauLifshitz}
	L.D. Landau and E.M. Lifshitz.
	\newblock {\em The Classical Theory of Fields. Course of Theoretical Physics.},
	volume~2.
	\newblock Butterworth-Heinemann, 1995.
	
	\bibitem[LLP84]{LandauLifshitz84}
	L.D. Landau, E.M Lifshitz, and L.P. Pitaevskii.
	\newblock {\em Electrodynamics of Continuous Media}, volume~8 of {\em Landau
		and Lifshtz Course of Theoretical Physics}.
	\newblock Elsevier, 2 edition, 1984.
	
	\bibitem[Mau80]{Maugin80}
	G.A. Maugin.
	\newblock The method of virtual power in continuum mechanics: Application to
	coupled fields.
	\newblock {\em Acta Mechanica}, 35:1--70, 1980.
	
	\bibitem[PH67]{Penfield_Haus}
	P.~Penfield and H.A. Haus.
	\newblock {\em Electrodynamics of Moving Media}.
	\newblock M.I.T. Press, 1967.
	
	\bibitem[PP62]{Panofsky1962}
	W.K.H. Panofsky and M.~Phillips.
	\newblock {\em Classical Electricity and Magnetism}.
	\newblock Addison-Wesley, 2nd edition, 1962.
	
	\bibitem[RdL05]{Raab_Lange}
	R.E. Raab and O.L. de~Lange.
	\newblock {\em Multipole Theory in Electromagnetism: Classical, Quantum, and
		Symmetry Aspects, with Applications}.
	\newblock Oxford University Press, 2005.
	
	\bibitem[Sch66]{Schwartz1966-short}
	Laurent Schwartz.
	\newblock {\em {Th\'eorie des distributions}}.
	\newblock Hermann, 1966.
	
	\bibitem[Seg23]{Segev_Book_2023}
	R.~Segev.
	\newblock {\em Foundations of Geometric Continuum Mechanics}.
	\newblock Birkhauser, 2023.
	
	\bibitem[Str41]{Stratton41}
	J.A. Stratton.
	\newblock {\em Electromagnetic Theory}.
	\newblock McGraw-Hill, 1941.
	
	\bibitem[TdCMG04]{Torres_etc_Multipoles}
	G.~F. Torres~del Castillo and A.~Mendez-Garrido.
	\newblock Differential representation of multipole fields.
	\newblock {\em JOURNAL OF PHYSICS A}, 37:1437--1441, 2004.
	
	\bibitem[Zan13]{Zangwill}
	A.~Zangwill.
	\newblock {\em Modern Electrodynamics}.
	\newblock Cambridge University Press, 2013.
	
\end{thebibliography}

\end{document}